\documentclass{elsart}

\usepackage{natbib}

\usepackage{graphicx}

\usepackage{epstopdf}

\usepackage{amssymb}

\begin{document}

\begin{frontmatter}



\title{The Cosmic Stellar Birth and Death Rates}


\author{John F. Beacom}

\address{Department of Physics and Department of Astronomy,
The Ohio State University, Columbus, Ohio 43210, USA}

\begin{abstract}
The cosmic stellar birth rate can be measured by standard astronomical
techniques.  It can also be probed via the cosmic stellar death rate,
though until recently, this was much less precise.  However, recent
results based on measured supernova rates, and importantly, also on
the attendant diffuse fluxes of neutrinos and gamma rays, have become
competitive, and a concordant history of stellar birth and death is
emerging.  The neutrino flux from all past core-collapse supernovae,
while faint, is realistically within reach of detection in
Super-Kamiokande, and a useful limit has already been set.  I will
discuss predictions for this flux, the prospects for neutrino
detection, the implications for understanding core-collapse
supernovae, and a new limit on the contribution of type-Ia supernovae
to the diffuse gamma-ray background.
\end{abstract}

\begin{keyword}
cosmic star formation rate \sep supernovae \sep neutrino background
\sep gamma-ray background
\PACS 97.60.Bw \sep 98.70.Vc
\end{keyword}

\end{frontmatter}


\section{Introduction}

The initial state that leads to a type-II (or type-Ib or type-Ic)
supernova is essentially an iron white dwarf (the endpoint of nuclear
fusion reactions) in the center of a massive star; when it reaches the
Chandrasekhar mass, this core will collapse, along with the rest of
the star. The direct microphysical messengers of the gravitational
explosion are the neutrinos emitted from the hot and dense
proto-neutron star.  The detectable (few tens of MeV) neutrinos are
emitted on a timescale of several seconds, corresponding to their
diffusion outward.  These neutrinos from a specific type-II supernova
have only been detected once, from SN 1987A~\citep{Hirata,Bionta}.

The initial state that leads to a type-Ia supernova seems to be a
carbon-oxygen white dwarf (the endpoint of nuclear fusion reactions in
a low-mass star) that accretes material from a bound binary companion;
when it reaches the Chandrasekhar mass, it will experience runaway
nuclear burning.  The direct microphysical messengers of the
thermonuclear explosion are the gamma rays from the nuclear decays of
freshly produced $^{56}$Ni and $^{56}$Co.  The detectable (few MeV)
gamma rays are emitted on a timescale of up to several months,
corresponding to the nuclear lifetimes.  These gamma rays have never
been convincingly detected from a specific type-Ia supernova, though
there have been three cases in which restrictive limits were placed
(e.g.~\cite{Milne}).

The difficulty is that nearby supernovae are rare.  Although gamma
rays are of course easier to detect, type-Ia supernovae are more rare,
and the number of emitted particles per supernova is less.  In both
neutrino astrophysics and gamma-ray astrophysics, the detection of a
nearby supernova is a high priority.  (Proposed neutrino detectors
should be able to reach even to several Mpc, if neutrinos can be
detected one or two at a time using a coincidence technique
\citep{Ando}.)  It should be noted immediately that the optical
emission from both types of supernovae, while readily detected even at
great distances, does not directly or faithfully reveal the details of
the explosions or the attendant extreme conditions that allow new
tests of particle physics.

I probably don't need to further convince anyone that detecting these
direct messengers of supernova explosions would be good.  But how?
What I always say is ``{\bf Everyone complains about the supernova
rate, but nobody does anything about it.}"  While we can't wish up a
nearby supernova, we can attempt to detect the diffuse glows of
neutrinos and gamma rays made by all past type-II and type-Ia
supernovae, respectively (practically speaking, redshifts less than
about $z = 1$ are the most important for detection).  I will argue
that while this is very challenging, the signal is always there, and
the prospects are quite encouraging.  The parallel nature of the above
discussion now develops an interesting split.  The neutrino background
has never been detected, but it is believed to be dominantly produced
by type-II supernovae.  On the other hand, the gamma-ray background
has been detected, but it is now believed that the contribution from
type-Ia supernovae is subdominant.

\begin{figure}
\centerline{\includegraphics[width=2.85in]{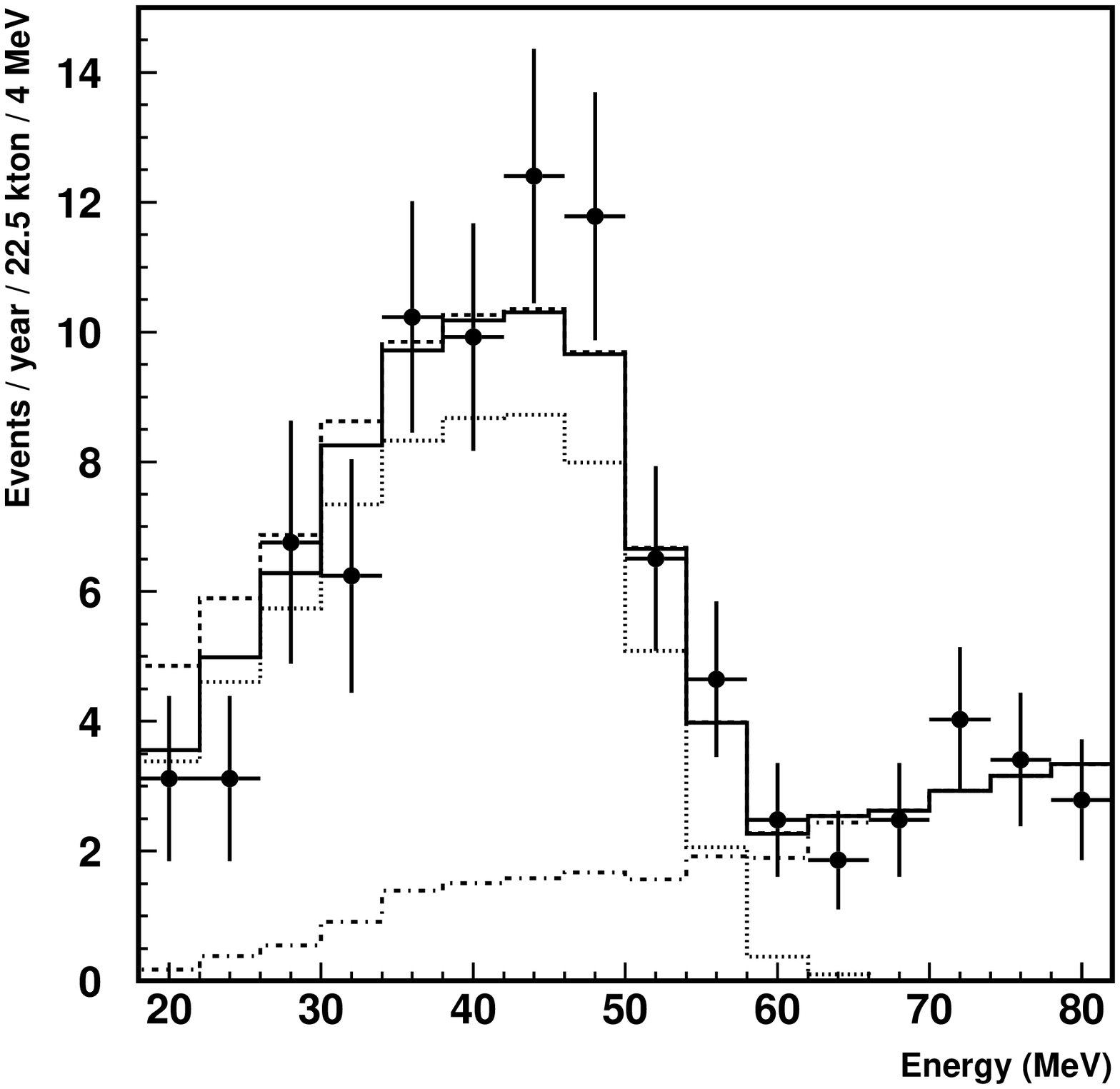}
\includegraphics[width=2.7in]{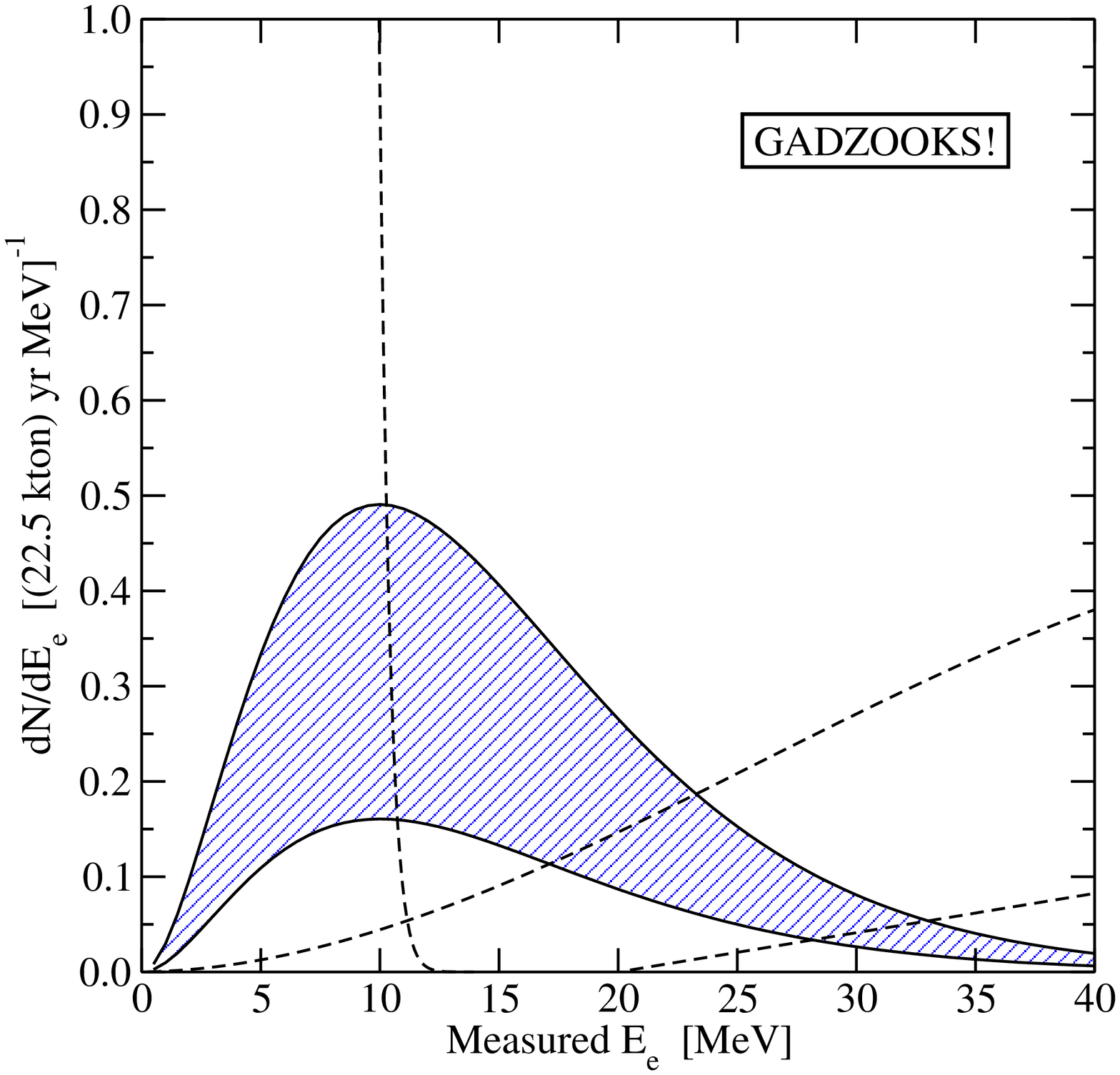}}
\caption{\label{fig:malek} {\bf Left figure:} The measured spectrum
(points with error bars) of $e^-$/$e^+$-like events in
Super-Kamiokande.  The solid line is the sum of the two atmospheric
neutrino background components, shown as lines below the solid line.
The dashed line {\it above} the solid line shows the largest allowed
fit for the detector background plus the type-II supernova neutrino
signal.  {\bf Right figure:} The projected spectrum if dissolved
gadolinium is added to Super-Kamiokande.  The shaded band is the
allowed range of supernova neutrino signals; our more recent
work~\citep{Concordance} favors the largest values.  The atmospheric
neutrino backgrounds have here been reduced by a factor $\sim 5$,
which would allow a much lower threshold, more like 10 MeV, where the
background from reactor neutrinos becomes overwhelming.  In the figure
on the right, the counts are per 1-MeV bin, whereas on the left, they
are per 4-MeV bin.  [Left figure from \cite{Malek}.  Right figure
adapted from \cite{GADZOOKS!}; the arXiv version defines
``GADZOOKS!".]}
\end{figure}

\cite{ClaytonSilk} proposed long ago that the prospects for detecting
the gamma-ray background from type-Ia supernovae were quite promising.
It wasn't until about 15 years later that people even started
considering the neutrino background from type-II supernovae; for early
work, see \cite{Seidov}, \cite{KGS}, and \cite{Woosley}.  Until very
recently, the prospects for detecting the neutrino background from
type-II supernova were thought to be at best exceedingly unlikely.
However, in 2003, the Super-Kamiokande collaboration published a flux
limit~\citep{Malek} that was about 200 times more restrictive than the
earlier limit from Kamiokande~\citep{Zhang}, and in fact in the range
of realistic models.  One surprising point that I will argue is that
this limit on the neutrino background strongly constrains the star
formation rate;\ with reasonable assumptions, this then constrains the
type-Ia supernova rate.  Using this limit, or going directly from the
measured type-Ia supernova rate data, one finds that the corresponding
gamma-ray background is well below the measured data.  Thus it now
appears {\it easier} to detect the neutrino background from type-II
supernovae than the gamma-ray background from type-Ia supernovae,
which is certainly not what anyone expected.

I will take the Super-Kamiokande neutrino limit as a starting point,
and then briefly review these constraints,
following~\cite{Concordance}.  The prospects for very significantly
improving the Super-Kamiokande sensitivity are discussed
in~\cite{GADZOOKS!}.  In brief, it was proposed that adding about
0.2\% dissolved gadolinium trichloride to Super-Kamiokande would allow
the detection of neutrons by their radiative captures on gadolinium.
This would then allow a coincidence detection of the positron and the
neutron in $\bar{\nu}_e + p \rightarrow e^+ + n$, which would greatly
reduce backgrounds which can mimic the positron by itself.  This
proposal is undergoing extensive research and development testing by
Vagins, and the prospects continue to be quite promising.

\begin{figure}
\centerline{\includegraphics[width=4in]{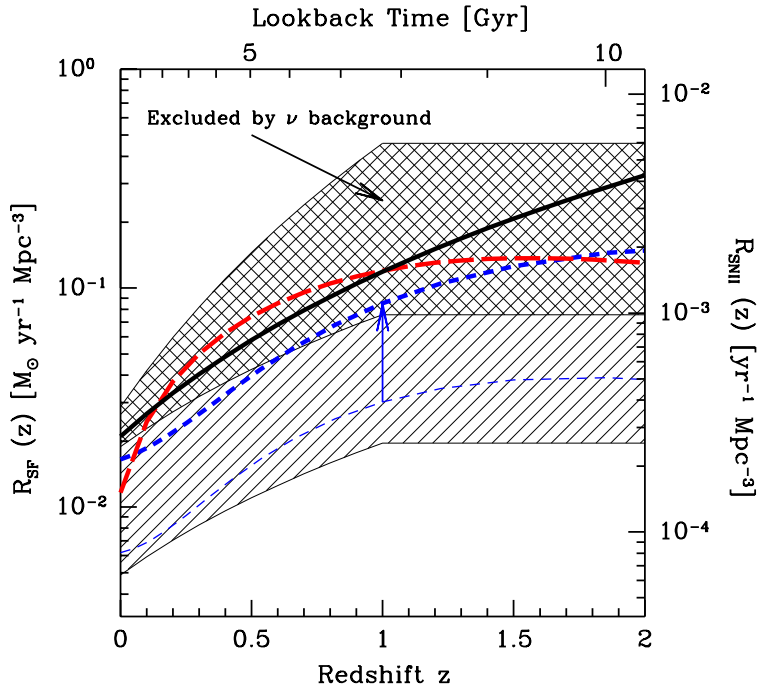}}
\caption{\label{fig:models} Recent measurements of the cosmic star
formation rate history.  The entire shaded region is consistent with
the results of the 2dF and SDSS cosmic optical spectrum.  The upper
cross-hatched region is ruled out by the limit on the type-II
supernova neutrino flux, while the lower shaded region is allowed.
Three recent (dust corrected by those authors) results are also shown
in the middle of the region; in one case, the result without dust
correction is shown for illustration.  The heavy solid line is the
result from GALEX~\citep{GALEX}.  The concordance region is defined by
the tension between the neutrino bound and the astronomical data, and
is thus concentrated at the upper edge of the lower band.  [Figure
taken from \cite{Concordance}.]}
\end{figure}


\section{The Cosmic Stellar Birth and Death Rates}

Recently, the measurements of the star formation rate history have
improved dramatically (e.g., see the summary in~\cite{Hopkins}).  An
important aspect of this is the larger and better-understood
corrections needed to compensate for obscuration by dust.  As a very
recent example of a new measurement, the GALEX results~\citep{GALEX}
are especially noteworthy.  Using the new Super-Kamiokande neutrino
flux limit~\citep{Malek}, an upper limit can be placed on the
normalization of the star formation rate history.  This requires
making some reasonable assumptions about the neutrino emission per
supernova.

\begin{figure}
\centerline{\includegraphics[width=3.5in]{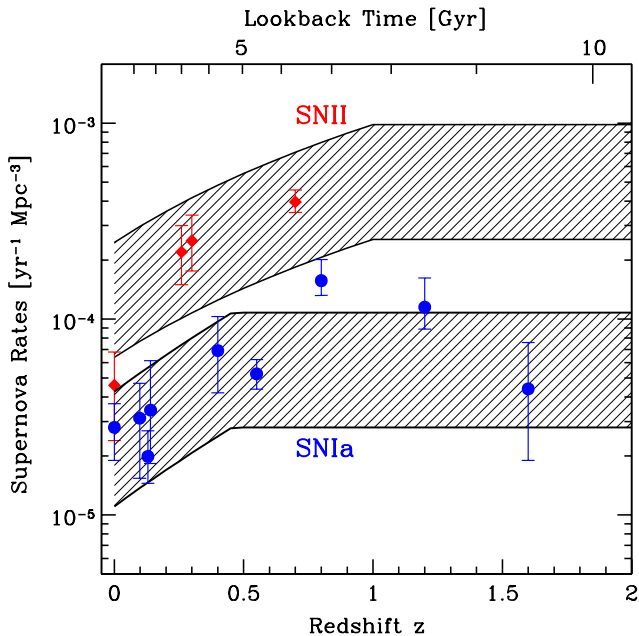}}
\caption{\label{fig:SNrates} The corresponding ranges for the type-II
and type-Ia supernova rates, following from the allowed lower band
from the previous figure.  [Figure taken from \cite{Concordance}.]}
\end{figure}

In \cite{Concordance}, we showed how the neutrino flux limit
significantly constrains the allowed range of realistic star formation
rate histories.  Interestingly, the dust-corrected results from GALEX,
for example, come in just above our deduced limit.  Very likely, this
means that the neutrino emission per supernova is somewhat less than
assumed.  On the other hand, it also means that the true flux is
probably close to the Super-Kamiokande limit, indicating that with
improved sensitivity~\citep{GADZOOKS!}, a discovery could soon be
made.  These results for the neutrino background are similar to other
recent calculations, e.g.,
\cite{Fukugita,Strigari,AndoSato,Lunardini,Daigne}, up to some
variations in the chosen inputs.  Since the astronomical factors are
already well known (and their precision is rapidly improving), a
measurement of the supernova neutrino background would alternatively
allow a new direct measurement of the supernova neutrino emission
parameters~\citep{Yuksel}, which would help resolve the lingering
mysteries of the SN 1987A data (e.g., as displayed by~\cite{Mirizzi}).

We defined a ``Concordance Model" by the tension between the upper
limit from the neutrino data, and the lower limit from the
astronomical data (i.e., these results with a somewhat smaller
correction for dust).  To validate this model, we considered the
type-II and type-Ia supernova rate histories.  Matching the type-II
supernova rate history depends just on the assumed stellar initial
mass function, whereas matching the type-Ia supernova rate history
also requires making some reasonable assumptions about the formation
efficiency and typical time delay between formation and
explosion~\citep{Watanabe}.  Using these results, which were also
confirmed by other indicators, we predicted the gamma-ray background
from type-Ia supernovae, showing that it must be far below the
measured data, which is surprising (see also similar results from
\cite{Ahn}).

\begin{figure}
\centerline{\includegraphics[width=3.5in]{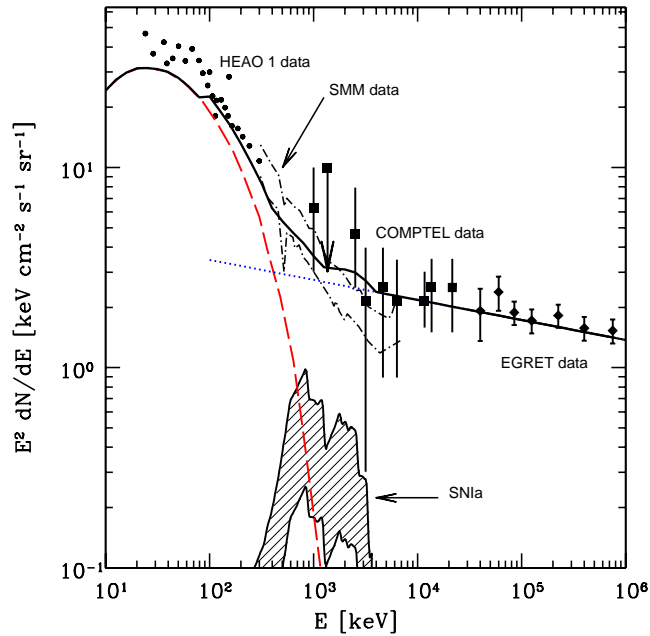}}
\caption{\label{fig:MeVSpectrum} Using the previous results, the
prediction for the type-Ia supernova gamma-ray background; the shaded
band has the same meaning as the previous figures.  [Figure taken from
\cite{Concordance}.]}
\end{figure}


\section{Conclusions}

Why would it be so important to detect the diffuse glows of neutrinos
and gamma rays from past supernovae?  That depends on your
perspective, but I should begin by reminding you that, except for the
$\simeq 20$ neutrinos from SN 1987A, and the at-most weak hints for
the gamma rays from $^{56}$Ni/$^{56}$Co from three nearby type-Ia
supernovae, these emissions have never been directly detected.  That's
strange, and frustrating, since we know that they {\it must} be
produced, and that they arise from the same physics that sources the
explosions.  For type-II supernovae, the neutrino emission can be
directly estimated from the final mass and radius of the neutron star,
which determines the gravitational energy release of $\sim 10^{53}$
erg.  For type-Ia supernovae, the gamma ray emission can be directly
estimated from the amount of synthesized $^{56}$Ni, which decays to
$^{56}$Co and then stable $^{56}$Fe.  These decays also power the
light curve, and the deduced thermonuclear energy release is $\sim
10^{51}$ erg.

For the neutrino background, the existing limit is very constraining,
allowing only something close to the most recent predictions for the
contribution from type-II supernovae.  If any other source contributed
significantly, it would thus be a big surprise; the present upper
limit on the flux of course applies also to any new sources.  Once the
neutrino flux is discovered, the spectrum shape can be tested, and the
shape corresponding to type-II supernovae is quite distinctive.  The
primary test of new physics is thus connected with the received
neutrino flux per type-II supernova.  This may be affected by neutrino
mixing among the known active flavors, possible mixing with postulated
sterile neutrinos, neutrino decay en route, modified neutrino emission
because of significant emission of new particles, etc.  The neutrino
emission per supernova can also be tested by effects on
nucleosynthetic yields, e.g., \cite{Yoshida}.

For the gamma-ray background, the situation is reversed, in that the
observed background is a more sensitive test of new physics than the
gamma ray emission per type-Ia supernova.  There are measurements of
the gamma-ray background, albeit with large error bars, over a broad
range of energies.  Just judging by the spectrum shape, there is no
clear indication of the characteristic shape corresponding to the
contribution from type-Ia supernovae near 1 MeV, and there are clearly
other sources at lower and higher energies.  An even stronger limit on
the contribution from type-Ia supernovae can be set using the star
formation and type-Ia supernova rates.  The properties of photons are
known, and it is believed that the physical conditions of the
explosion do not generally permit significant perturbations due to new
physics.  However, models of new physics, including the decay or
annihilation of dark matter (or other particles arising in
extra-dimensional models), can contribute significantly to the
observed gamma-ray background.  Once the measurements improve, and the
astrophysical components are better understood, they can be
subtracted, leading to more stringent limits on the contributions due
to new physics.  If the type-Ia supernova contribution can somehow be
isolated~\citep{CGB}, it will provide a new test of the gamma-ray
emission per supernova and the evolution of the supernova rate, both
of which could help improve the understanding required for using
type-Ia supernovae as standard candles to measure dark energy.

To summarize, a better understanding of the diffuse neutrino and
gamma-ray backgrounds, and specifically the portions associated with
supernovae, would have important implications for several fields:
\begin{itemize}
\item {\bf Particle physics:}
Neutrino properties; novel energy-loss channels in type-II supernovae; etc.
\item {\bf Nuclear physics:} Production of the light and heavy
elements; neutron star equation of state; etc.
\item {\bf Astrophysics:} Cycle of stellar birth, life, and death;
constraints on new sources; etc.
\item {\bf Cosmology:} Supernova distance indicators and dark energy;
dark matter decay or annihilation; etc.
\end{itemize}
Thus the faint fossil records of supernovae, revealed by their diffuse
neutrino and gamma ray backgrounds, offer a challenging but very
promising route to better understanding the physics of their
explosions.

This work was supported by The Ohio State University.  I thank the
organizers for the invitation to this very nice meeting, and also
especially Eli Dwek, Peter H\"oflich, and Stan Woosley for their
interesting and helpful questions and comments on my talk.  I thank
Shin'ichiro Ando, Louie Strigari, Mark Vagins, Terry Walker, Hasan
Y\"uksel, and Pengjie Zhang for very enjoyable collaborations on this
and closely related work.


\appendix



\end{document}